\documentstyle[eqsecnum,prb,aps,twocolumn,epsf]{revtex}

\def\be{\begin{equation}}
\def\ee{\end{equation}}

\begin{document}

\draft

\title{~\\~\\ Numerical study of current distributions\\
in finite planar Hall samples with disorder}
 \author{K. Shizuya}
  \address{Yukawa Institute for Theoretical Physics\\
 Kyoto University,~Kyoto 606-8502,~Japan}

\maketitle
 
\begin{abstract} 
A numerical study is made of current distributions in finite-width
Hall bars with disorder and some theoretical observations are
verified. The equilibrium current and the Hall current are
substantially different in distribution.
It is observed that in the Hall-plateau regime the Hall current tends
to concentrate near the sample edges while it diminishes on average 
in the sample interior as a consequence of localization.  
The edge states themselves scarcely affect both 
the equilibrium and Hall currents in the plateau regime. 
The sample edge combines with disorder to efficiently delocalize
electrons near the edge while the Hall field competes with disorder to
delocalize electrons in the sample interior. 
A possible mechanism is suggested for the breakdown
of the quantum Hall effect.
\end{abstract}


\section{Introduction}

The current distribution in a Hall sample is a subject of interest
pertaining to the foundation of
the quantum Hall
effect~\cite{V,Rev,AA,Pr,L,H,TK,LLP,MS,SKM,B,CSG,Wen,KS} (QHE).
The standard explanations of the QHE are based on a picture that 
the Hall current is carried by a finite fraction of electron states in
the sample bulk that remains extended in the presence of disorder;
there localization is essential~\cite{AA,Pr,L} 
for the formation of visible conductance plateaus.
An alternative explanation is based on a picture~\cite{MS,SKM,B,CSG}
in which the electron edge states~\cite{H} are taken to be principal
carriers of current, and leads to the concept of edge channels which
turns out useful in summarizing observed results.~\cite{WFSK} 
Still it is not quite clear experimentally whether 
the current flows only along the sample edges;
some experiments,~\cite{FKHBW,HTS,W} though indirectly, 
suggest edge-channel widths much
larger than the width naively expected from the edge-state picture of
the QHE. It is important to clarify how these apparently
different pictures could be related.~\cite{T,A,KStwo} Theoretical
consideration on current or potential distributions has so far
focused on clean samples~\cite{MRB} rather than disordered
samples.~\cite{OO}

The purpose of this paper is to present a numerical study of current
distributions in small Hall bars with disorder and to verify some
theoretical observations.  
We shall handle samples which accommodate as many 
as 280 electrons; this makes it possible to disclose features of 
the current distributions not observed 
in earlier numerical work~\cite{OO} on smaller samples.
In particular, we shall find evidence that in the Hall-plateau
regime the Hall current tends to concentrate near the sample edges
while it diminishes on average in the sample interior as a
consequence of localization.  
The edge states themselves scarcely affect the distributions of both 
the equilibrium and Hall currents in the plateau regime.

We shall use a uniform Hall field to detect the current-carrying
characteristics of Hall electrons.  One certainly has to treat 
the problem of an electrostatic potential selfconsistently, in which
charges will be redistributed as pointed out by Chklovskii et
al.~\cite{CSG}  What we propose here, however, is that, even before
the self-consistent potential is used, there exists an important
physics of redistribution of the Hall current via disorder, and 
we believe that even after the inclusion of the selfconsistency  
the essential feature will remain because the key mechanism is
competition of the electrostatic potential with disorder rather than
its distribution, as we shall see.

In Sec.~II we present a theoretical framework for our numerical
simulations.
In Sec.~III we discuss the influence of isolated impurities on
electron states.
We handle a Hall sample with an array of impurities 
in Sec.~IV and samples with random impurities in Sec.~V.
Section~VI is devoted to a summary and discussion.

\setcounter{section}{1}

\section{Formalism}

Consider electrons confined to a strip of length $L_{x}$ and width
$L_{y}$ (or formally, a strip bent into a loop of circumference
$L_{x}$), described by the Hamiltonian:
\begin{eqnarray}
&&{\sf H} =H_{0}+ U(x,y)-eA_{0}(y), \label{HAU}\\
&&H_{0} = {1\over 2}\omega \Bigl\{ \ell^{2}p_{y}^{2}
+(1/\ell^{2})(y - y_{0})^{2}\Bigr\},
 \label{hzero}
\end{eqnarray}
written in terms of $\omega \equiv eB/m$,
the magnetic length ${\ell}\equiv 1/\sqrt{eB}$ and 
$y_{0} \equiv p_{x}/(eB)$. 
Here the  Landau-gauge vector potential $(-By,0)$ has been used
to supply a uniform magnetic field $B$ normal to the plane.
We take explicit account of the edge $y=0$, where the wave function is
bound to vanish; the prescription for the other edge $y = L_{y}$ will
be specified later.

We shall detect the current $j_{x}(x,y)$ flowing in the presence of a
Hall potential  $A_{0}(y)$ and an impurity potential of the form
\begin{equation}
U(x,y) = \sum_{i}\lambda_{i}\, \delta (x - x_{i}) \delta (y - y_{i}),
 \label{Uxy}
\end{equation}
which consists of short-range impurities of strength $\lambda_{i}$
distributed over the sample at position $(x_{i},y_{i})$.
For simplicity, we here take $A_{0}(y)= - y E_{y}$ which  
produces a uniform Hall field $E_{y}$ in the $y$ direction.

The kinetic term $H_{0}$ describes the cyclotron motion of 
an electron, which gives rise to Landau levels 
$|N\rangle = |n, y_{0}\rangle$ labeled 
by integers $n = 0,1,2,\cdots$, and $y_{0}=\ell^{2}\,p_{x}$,
with wave functions of the form
\begin{equation}
\langle x,y |N\rangle = 
(L_{x})^{-1/2}\,e^{ixp_{x}}  \phi_{n} (y; y_{0}). \label{psiN}
\end{equation}
The $\phi_{n} (y; y_{0})$ are given~\cite{MS} by the parabolic
cylinder functions~\cite{WW}  $D_{\nu}(\sqrt{2}(y-y_{0})/\ell)$, which  
for electrons residing in the sample bulk are reduced to the
usual harmonic-oscillator wave functions.
Setting $\phi_{n}=0$ at $y=0$ fixes the energy eigenvalues
\begin{equation}
\epsilon_{N}= \omega \{{\nu_{n}}(y_{0})+ {\textstyle {1\over{2}}} \}
\label{eN} 
\end{equation}
as a function of $y_{0}$ for each $n$.
The spectra $\nu_{n}(y_{0})$ rise sharply only  
for $y_{0} \lesssim O(\ell)$ and recover integer values $n$ 
for $y_{0}\gtrsim O($a~few~$\ell$) ;~\cite{MS}
for the $n=0$ level, $\nu_{0}(y_{0})$ decreases 
from 1 to 0.003 as $y_{0}$ varies from $0$ to $2.5 \ell$.

The presence of the sample edge $y=0$ has effectively generated 
strong potentials $\omega \nu_{n}(y_{0})$ that confine electrons
to a region $y_{0}\gtrsim 0$ in $y_{0}$ space and 
that drive electrons along the edge.   
Indeed, near the sample edge an electron state $(n,y_{0})$ 
travels with velocity 
$v_{x}\approx \omega \ell^{2} {\nu_{n}}'(y_{0}) \sim - \omega \ell$
much larger than the field-induced drift velocity $E_{y}/B$.
Classically these edge states~\cite{H,MS,SKM,B,Wen} are
visualized as electrons hopping along the sample edge and
carry current even in equilibrium.

The $\phi_{n} (y; y_{0})$ are highly localized around 
$y\sim y^{\rm cm}\equiv y_{0}-\ell^{2}{\nu_{n}}'(y_{0})$ 
with spread $\triangle y \sim O(\ell)$; see Eq.~(\ref{yNN}) below.
With $\langle x,y|N\rangle$ taken to be periodic in $x$,
$y_{0}=\ell^{2}p_{x}$  is labeled by integers $k$:
\begin{equation}
y_{0}= (2\pi\ell^{2}/L_{x}) k \equiv y_{0}[k]. 
\label{yzerok}
\end{equation}
We shall henceforth use this discrete label $k$
rather than $y_{0}$ to refer directly to each electron state,
denote $|N\rangle =|n, y_{0}[k] \rangle$ simply as $|n, k \rangle$
and use the normalization 
$\langle N|N'\rangle = \delta_{nn'}\delta_{kk'}$.

For numerical calculations it is advantageous to handle the
Hamiltonian $H_{NN'}\equiv \langle N|{\sf H}|N'\rangle$ 
in $N=(n,k)$ space rather than in real $(x,y)$ space:
\begin{eqnarray}
H_{N N'}\!
&=&  \omega \{ \nu_{n}(y_{0})\!+{\textstyle {1\over{2}}} \}\,
\delta_{NN'}\!+ U_{NN'}\! +eE_{y}\, y_{NN'},
\label{HNN}   
\end{eqnarray}
with
\begin{eqnarray}
U_{NN'}\!&=&\! {1\over{L_{x}}} 
\sum_{i} \lambda_{i} e^{-i(y_{0}\! - y'_{0}) x_{i}/\ell^{2}}\! 
\phi_{n}(y_{i};\!y_{0}) \phi_{n'}(y_{i};\!y'_{0}),
\label{UNN} \\
y_{NN'}&=& \{y_{0}\,\delta_{nn'} 
+ \ell\,Y_{n n'}(y_{0}) \}\, \delta_{k k'}  , 
\label{yNN} 
\end{eqnarray}
where $U_{NN'}\equiv \langle N|U|N'\rangle,\  
y_{NN'} \equiv \langle N|y|N'\rangle, 
y_{0}=y_{0}[k]$ and $y'_{0}=y_{0}[k']$.
The matrix elements $Y_{n n'}(y_{0})$ 
are completely known~\cite{KStwo} through the spectra
$\nu_{n}(y_{0})$:
\begin{eqnarray}
Y_{nm}&=& (-1)^{n+m}\, \ell\,  
|{\nu_{n}}'{\nu_{m}}'|^{1/2}
 /[1-(\nu_{n} -\nu_{m})^{2}]  \label{ynn}  
\end{eqnarray}
with 
${\nu_{n}}'= \partial_{y_{0}}\nu_{n}(y_{0})$; 
in particular, $Y_{n n}(y_{0}) = - \ell\, {\nu_{n}}'(y_{0})$.
For $y_{0} \gg O(\ell)$,  $Y_{nn'}$ are reduced to
constant matrices  
$Y_{nn'}^{(\rm bulk)}= \sqrt{n'/2}\, \delta_{n+1,n'} 
+ \sqrt{n/2}\, \delta_{n,n'+1}$.

When the disorder $U_{NN'}$ is weak compared with the level gap,
i.e., $|\lambda_{i}/2\pi\ell^{2}| < \omega$, one can diagonalize 
$H_{NN'}$ with respect to Landau-level labels $(n,n')$ 
by a suitable unitary transformation $H^{W}= WHW^{-1}$ 
so that the diagonal component 
$\langle N|H^{W}|N'\rangle =\delta_{nn'}H^{W}_{nn}(k,k')$ 
describes each impurity-broadened Landau subband.
In particular, with a simple $O(U)$ choice $W=e^{i\Lambda}$ with
\begin{equation}
i\Lambda_{NN'}= {e E_{y}\,y_{NN'}+ U_{NN'}
\over{\omega (\nu_{n}[k] - \nu_{n'}[k']) }}  \ \ \ \ 
({\rm for\ }n\not=n') ,
 \label{lamNN}
\end{equation}
the subband Hamiltonian $H^{W}_{nn}(k,k')$ reads 
\begin{eqnarray}
H^{W}_{nn}(k,k')
=&&\! \Bigl\{ \omega (\nu_{n}[k]\!+{\textstyle {1\over{2}}} ) + 
eE_{y} \{y_{0}\!-\!\ell^{2} {\nu_{n}}'(y_{0})\} \Bigr\}\,  
\delta_{k k'} \nonumber \\
&& + U_{nn}(k,k') + O(U^{2}/\omega),
\label{HWnn}
\end{eqnarray}
where we have set
$\nu_{n}(y_{0}[k])\rightarrow \nu_{n}[k]$ and $U_{NN'}\rightarrow 
U_{nn'}(k,k')$.

Some care is needed here. 
The choice~(\ref{lamNN}) becomes inadequate for 
$\nu_{n} - \nu_{n'} \sim 0$; such degeneracy among
different subbands is inherent to edge states.
Fortunately, to the order we work (i.e., $O(U)$), 
there is a practical way out of such complications:~\cite{fn}
In what follows we simply focus on the $n=0$ subband and 
consider only electron states with 
$y_{0}= (2\pi/L_{x}) k >0$ (i.e., $k = 1,2,\cdots$) that are free
from intersubband degeneracy.

The Hamiltonian governing the $n=0$ subband of our interest is
$H^{W}_{00}(k,k')$ with $k,k'=1,2,\cdots,N_{\rm max}$,
where $N_{\rm max}$ is the maximum number of electrons 
to be retained in numerical simulations.
Associated with each eigenstate $|\alpha)$ of $H^{W}_{00}$ is an 
($N_{\rm max}$-component) wave function 
$\alpha_{k}= \langle 0,k|\alpha)$,
which is found as an eigenvector by diagonalizing
$H^{W}_{00}(k,k')$ numerically.

The ${\bf x}$-space wave~functions  
$\langle {\bf x}|W^{-1}|\alpha)$
are readily reconstructed from these $\alpha_{k}$:
\begin{equation}
\langle {\bf x}| W^{-1}|\alpha) 
= \sum_{n,k}	\langle {\bf x}|n,k\rangle 
\left(\delta_{0n}\,\alpha_{k} +\beta^{\,n}_{\,k} \right) +\cdots,
 \label{xWalpha}
\end{equation}
where $\langle {\bf x}| \equiv \langle x,y|$ for short and 
\begin{equation}
\beta_{\,k}^{\,n}=
-{1\over{\omega}}\,\sum_{k'} {e\ell E_{y}Y_{n0}(y_{0}) \delta_{kk'}
+  U_{n0}(k,k')\over{\nu_{n}[k]-\nu_{0}[k']}}\, \alpha_{k'}
 \label{betank}
\end{equation}
for $n\ge 1$ and $\beta_{\,k}^{\,0}=0$.
The spatial distribution of the mode $\alpha$ is given by
\begin{equation}
\rho^{(\alpha)}(x,y)=|\langle  {\bf x} | W^{-1}|\alpha)|^{2}
 \label{rhoalpha}
\end{equation}
and the current density $j_{x}=e\omega\, \Psi^{\dag}
(y-{1\over2}\,\ell^{2}\!\stackrel{\leftrightarrow}{p}_{x})\Psi$ by
\begin{equation}
j_{x}^{(\alpha)}(x,y) = 
e\omega\, {\rm Re} \left[ (\alpha|W| {\bf x}\rangle 
\langle {\bf x}| (y-y_{0}) W^{-1}|\alpha) \right],
 \label{jxalpha}
\end{equation}
where $\langle {\bf x} |(y-y_{0}) W^{-1}|\alpha)$ is given by
Eq.~(\ref{xWalpha}) with $\langle {\bf x}|n,k\rangle$ replaced by 
$(y - y_{0}[k])\langle {\bf x}|n,k\rangle$. 
To extract the Hall-current component out of $j_{x}^{(\alpha)}$
numerically, we shall vary $E_{y}$ by a small amount~\cite{Wagner}
$\delta E_{y}$ and calculate the change  
\begin{equation}
j_{\rm Hall}^{(\alpha)}(x,y) = 
j_{x}^{(\alpha)}(x,y|E_{y}\!+\!\delta E_{y}) 
- j_{x}^{(\alpha)}(x,y|E_{y}) .
 \label{jhallalpha}
\end{equation}

For the simulations to be given in Secs.~III and IV 
we have constructed the normalized wave functions 
$\phi_{n}(y;y_{0})= C_{n}( y_{0}) D_{\nu}(\sqrt{2}(y-y_{0})/\ell)$
with $\nu \rightarrow \nu_{n}(y_{0})$ as real functions of $y$ 
and $y_{0}$ by determining the spectra $\nu_{n}(y_{0})$ 
and overall normalization $C_{n}( y_{0})$ numerically
to accuracy of 10 digits for $y_{0}\ge 0$ and $n=$0,1, and 2.
For each given configuration of impurities we shall diagonalize 
$H^{W}_{00}(k,k')$ of Eq.~(\ref{HWnn}) to $O(U)$, construct 
the ${\bf x}$-space wave functions~(\ref{xWalpha})
through the eigenvector $\alpha_{k}$,
and calculate the current $j_{x}^{(\alpha)}$ of Eq.~(\ref{jxalpha}).
For the wave functions~(\ref{xWalpha}) 
we shall take only the $n=0 \leftrightarrow n=1$ transitions 
($\beta^{1}_{k}$) into account; actually we have in some cases 
included the $n=0 \leftrightarrow n=2$ transitions as well and 
found no appreciable change in current distributions.

\section{Isolated impurities}

In the absence of impurities all electron states
$|n,y_{0}\rangle$ are plane waves extended in $x$, 
though localized in $y$ with spread $\triangle y \sim O(\ell)$.
Such spatial characteristics of electron states are modified in the
presence of disorder, and electrons
tend to be confined in finite domains of space (or trapped by impurities).
In this section we examine the influence of isolated impurities on
electron states.

For most of the cases to be discussed in this paper 
we set $e\ell E_{y}/\omega = - 1/10^5$, which
corresponds to a Hall field of strength $E_{y}\approx$ - 0.1 V/cm 
under a typical setting $\ell \sim$ 100 \AA\, 
and $\omega \sim 10$ meV.
[The choice $E_{y} <0$ is taken simply for convenience 
so that the $y=0$ edge is the ``upper'' edge.
We have examined the $E_{y} >0$ cases as well and 
found no essential change in our conclusion.]
To calculate $j_{\rm Hall}^{(\alpha)}$ of
Eq.~(\ref{jhallalpha})  we set $\delta E_{y}/E_{y}= 1/100$ and 
establish one-to-one correspondence between the eigenfunctions of
the $\delta E_{y}=0$ and $\delta E_{y}\not=0$ cases by direct
examination of their overlap.
In what follows  we shall employ dimensionless impurity strengths
$s_{i}$ defined by $\lambda_{i}/2\pi \ell^{2} = s_{i}\, \omega$, and 
measure energy relative to ${1\over2}\omega$.

Consider a Hall sample of length 
$L_{x}=\sqrt{2\pi}\ell \times 10 \approx 25 \ell$ and width
$L_{y}=\sqrt{2\pi}\ell \times 9 \approx 22.5 \ell$,
supporting 90 electron states in the lowest level $n=0$ with 
$0<y_{0}\le L_{y}$, or $y_{0}=(2\pi\ell^{2}/L_{x}) k$ with 
$k=1,2,\cdots, 90$.
The sample edge $y=0$ is a sharp edge while the region $y \sim L_{y}$ 
is left impurity-free.

Let us place three impurities of strength $(s_{1},s_{2},s_{3}) = 
(0.1, 0.1, -0.1)$ at $y = (2.5\ell,10\ell,18\ell)$ 
along the line $x = 20\ell$ on the sample
and two more impurities of strength $(s_{4},s_{5}) = 
(0.05, -0.05)$ at $y = (8\ell,17\ell)$ along $x=10\ell$.
These impurities are sufficiently far apart from each other and can
practically be regarded as isolated impurities.
Four of them capture electrons, leading to four localized states of 
energy $\epsilon_{i}\approx s_{i}\, \omega$ with $i=2,3,4,5$,
as seen from the density distribution 
$\rho^{(\alpha)}(x,y)$ in Fig~\ref{fig1} (a).

To get a definite idea of localized states it is useful 
to refer to the case of an isolated impurity of strength 
$s = \lambda/(2\pi \ell^{2}\omega)$ placed at $x=y=0$ on an 
infinite plane with $E_{y}=0$.
In the $n=0$ level one localized state of energy 
$\epsilon \approx s\omega$
arises with a wave function of the form
\begin{equation}
\psi^{\rm loc}
= {1\over{\sqrt{2\pi\ell^{2}}}} 
e^{-{1\over{2\ell^{2}} }(r^{2} -2i xy)} 
\left\{ 1 - s\, (1 - {r^{2}\over{2\ell^{2}}})\right\},
		\label{locwf}
\end{equation}
where $r^{2}=x^{2}+y^{2}$; the rest of the $n=0$ states have 
no energy shift, $\epsilon =0$.
The $O(s)$ correction comes from 
the $\beta^{1}_{k} \propto U_{10}$ term of Eq.~(\ref{betank}).
Numerically Eq.~(\ref{locwf}) accounts for the four localized states
in Fig.~\ref{fig1}~(a) very well. 
A closer look at Fig.~\ref{fig1} (a) reveals that a localized state
gets somewhat narrower for an attractive impurity $(s < 0)$;
thus attractive impurities are more efficient in trapping electrons.

A Hall field causes a drift of an orbiting electron and works to 
delocalize it.  A simple estimate of energy cost suggests that 
an isolated impurity of strength $s$ would fail to capture an electron
for
\begin{equation}
|s|\,\omega = |\lambda|/(2\pi \ell^{2}) \lesssim e \ell\, |E_{y}| .
  	\label{svsE}
\end{equation}  
Indeed, we have verified that an impurity as weak as $s = 10^{-4}$ 
can capture an electron while it fails for $s = 10^{-5}$.

Similarly, a confining potential $\omega \nu_{0}(y_{0})$ works to 
delocalize electrons. The edge states driven by an ``effective" field
as strong as  
$e\ell E_{y}^{\rm eff} \sim \omega \ell\,{\nu_{0}}' \sim \omega$
are robust against disorder and remain extended.~\cite{H}  
Indeed, Fig.~\ref{fig1} (a) demonstrates explicitly that 
an impurity of strength $s_{1}=0.1$, situated (at $y=2.5\ell$) 
close to the edge, 
fails to trap any electron. (Note that 
$\ell\,{\nu_{0}}' \approx -0.011$ for $y_{0}=2.5\ell$.)

Figure~\ref{fig1} (b) shows the current distribution 
$j_{x}^{(\alpha)}(x,y)$, plotted in units of $-e\omega/L_{x}$, 
for each of the five states.
There each localized state is accompanied by a circulating current 
distribution that reflects the underlying counterclockwise 
cyclotron motion of an electron. 
These localized current distributions are not very sensitive to 
impurity strengths $s_{i}$.
In contrast, as seen from Fig.~\ref{fig1} (c),
the Hall current density $j_{\rm Hall}^{(\alpha)}(x,y)$ localized 
about an impurity exhibits characteristic local variations, 
which depend on the nature, attractive or repulsive, 
of the impurity and which, somewhat unexpectedly, get amplified in
magnitude and range as the impurity becomes weaker.

These features of $j_{x}^{(\alpha)}$ and $j_{\rm Hall}^{(\alpha)}$ are 
readily understood from the current density~\cite{OO} 
about the localized state~(\ref{locwf}): 
\begin{equation}
j_{x}^{\rm loc}(x,y) = {e\omega\over{4\pi \ell^{2}}} 
\left\{ y - {eE_{y}\over{s\,\omega}} (y^{2} - \ell^{2})
\right\}\, e^{-r^{2}/2\ell^{2}},
	\label{jxloc}
\end{equation}
where, for simplicity, the $O(E_{y})$ correction has been calculated 
by treating the Hall potential as a perturbation solely to the 
$O(s^{0})$ part of $\psi^{\rm loc}$.
Note that the net current vanishes, 
$\int dy\, j_{x}^{\rm loc}(x,y)=0$.

For a filled subband the effect of impurities disappears 
from the Hall-current distribution  
$j_{\rm Hall}(x,y) = \sum_{\alpha}j_{\rm Hall}^{(\alpha)}(x,y)$, 
which then recovers the same distribution, uniform in $x$, as in the
impurity-free case; this is verified by use of Eq.~(\ref{xWalpha}).
In view of this fact, a {\em vacant} localized state
about a repulsive $(\lambda>0)$ impurity and an {\em occupied} localized
state about an attractive $(\lambda<0)$ impurity act alike, 
as far as $j_{\rm Hall}(x,y)$ is concerned.  
For such a state 
$j_{\rm Hall}(x,y)$ flows on both sides of an impurity in the
same direction as if it avoids the impurity center [like the 
$j_{\rm Hall}^{(\alpha)}(x,y)$ about an attractive impurity 
in Fig~\ref{fig1}(c)]; 
here we already see the potential that each impurity expels the Hall current
out of its center.

\section{An array of impurities}

Let us next consider a Hall sample of length 
$L_{x}=\sqrt{2\pi}\ell \times 12 \approx 30 \ell$ and width
$L_{y}=\sqrt{2\pi}\ell \times 9 \approx 22.5 \ell$,
on which 63 impurities of equal strength $s=0.2$ are placed at
``even'' points $(x_{i},y_{j})=( 2(i+2)\ell, 2(j+1)\ell )$ with 
$i=1,2,\cdots,9$ and $j=1,2,\cdots,7$ within the domain 
$6\ell \le x \le 24 \ell$ and $4 \ell \le y \le 16 \ell$.
We have distributed impurities in a regular pattern so as to observe
the cooperative effects of many impurities in an amplified fashion.
The remaining region $16 \ell < y \le L_{y}$ is left
impurity-free so that it can simulate a gentle edge region supporting
extended electron states.

This sample supports 108 electron states in the lowest subband $n=0$ 
with $0<y_{0}\le L_{y}$.
We arrange them in the order of descending energy and use the
number of vacant states, $N_{\rm v}= 108 - N_{e}$, to specify
the filling of the subband.
For convenience we start with the filled subband $N_{\rm v}= 0$ 
and study the currents this subband supports, 
$j_{x}(x,y) = \sum_{\alpha} j_{x}^{(\alpha)}(x,y)$ and
$j_{\rm Hall}(x,y) = \sum_{\alpha} j_{\rm Hall}^{(\alpha)}(x,y)$,
by removing electrons one by one.
In Fig.~\ref{fig2} (a) we plot the total currents 
$J_{x} = \int dy\, j_{x}(x,y)$ and 
$J_{\rm Hall} = \int dy\, j_{\rm Hall}(x,y)$ as 
functions of $N_{\rm v}$,  
in units of $0.05\times(- e\omega \ell/L_{x})$ and 
$-(e \delta E_{y}/B)(1/L_{x})$, respectively.

Direct inspection of the spatial distribution $\rho^{(\alpha)}(x,y)$ 
for each state reveals that most of the first 48 states are localized,
except for the edge states of
energy $\epsilon_{1} \approx 0.78\,\omega, 
\epsilon_{2}\approx 0.59\,\omega,
\epsilon_{3}\approx 0.44\,\omega, 
\epsilon_{22}\approx 0.31\, \omega,$ etc.
The edge states are readily identified by a large amount of current
they carry.
Actually they are responsible for remarkable steps in $J_{x}$, 
which emerge each time a new edge state becomes vacant with increasing
$N_{\rm v}$. 
In contrast, they leave little effect on
$J_{\rm Hall}$ in Fig.~\ref{fig2} (a), where a Hall plateau
persists for $N_{\rm v} \lesssim 48$. 
In fact, the first three edge states combine to carry a larger amount 
of current  $j_{x}$ than the rest of states combined, but the amount 
of Hall current they carry is less than a single unit 
$(-e \delta E/B)(1/L_{x})$.

This leads to an important observation:
In the regime supporting clear Hall plateaus,
the fast-traveling edge states are practically vacant and scarcely
contribute to the currents $j_{x}(x,y)$ and 
$j_{\rm Hall}(x,y)$.
The edge states, traveling in the negative $x$ direction, 
give rise to a prominent edge current (of paramagnetic nature)
for the filled subband $(N_{\rm v}=0)$. 
However, as soon as the first three edge states become vacant, 
the orbital diamagnetic current takes over and $j_{x}(x,y)$ reverses
its direction in the edge region, as seen from the current profile 
across the sample width at $x=16\,\ell$ in Fig.~\ref{fig2} (b).
It is now the electron bulk states that are responsible 
for the prominent current distributions  
flowing in opposite directions at opposite edges for 
$4\lesssim N_{\rm v}\lesssim 50$.

As $N_{\rm v}$ is increased within the plateau regime, 
localized states disappear one by one from inside the disordered
domain, and those slightly above the mobility edge are 
seen to lie around its periphery.
When the disordered domain gets well depopulated, 
i.e., for $40 \le N_{\rm v} \lesssim 50$, 
$j_{x}(x,y)$ almost vanishes inside it and there
emerges a current circulating along its periphery, as seen clearly
around $y \sim 3\,\ell$ and $y \sim 17\,\ell$ in the $N_{\rm v}=45$
case of Fig.~\ref{fig2} (b).  
This current is a diamagnetic current associated with
the extended states surviving in the regions 
$0 \lesssim y \lesssim 5\,\ell$  and 
$16\,\ell \lesssim y \lesssim 22.5\,\ell$ of the sample,
clearly seen from the density profile 
$\rho(x,y)=\sum_{\alpha}\rho^{(\alpha)}(x,y)$ 
at $x=16\,\ell$.

In the plateau regime  $4\le N_{\rm v} \le 48$, 
the Hall current remains almost constant in net amount but 
changes in distribution with $N_{\rm v}$.
When the disordered domain becomes well depopulated, 
$j_{\rm Hall}(x,y)$ virtually vanishes inside it and  
flows along its periphery in the same direction, forming two localized
distributions that become prominent as $N_{\rm v}\rightarrow 48$. 
See the current distribution for $N_{\rm v} = 45$
and its profile at $x=16\,\ell$, shown in Fig.~\ref{fig3}. 
Here we see explicitly that the Hall current is expelled out of 
the disordered portion of the sample.
We have verified that even for smaller impurity strengths 
$s_{i}=0.1$ and $s_{i}=0.01$ the current distributions show
essentially the same characteristics.

\section{Randomly-distributed impurities}

In this section we examine Hall samples with random impurities.
One of the samples we have considered has length 
$L_{x}=\sqrt{2\pi}\ell \times 28 \approx 70 \ell$ and width
$L_{y}=\sqrt{2\pi}\ell \times 8 \approx 20 \ell$,
supporting 224 electron states 
in the $n=0$ subband with $0 \le y_{0} \le 20 \ell$.\break
180 short-range impurities of varying strength 
within the range $ |s_{i}| \le 0.1$ 
are randomly distributed over the full length and part of the
width, $0 < y \le 15 \ell$.  
The remaining region $15 \ell <  y \le 20 \ell$ 
is left impurity-free as before.
We have employed a number of impurity configurations for this sample
and obtained qualitatively the same conclusion.
Here we select one configuration which is generated by
reshuffling of a basic configuration of 45 impurities and which
therefore is readily written out; see the appendix for details.

Let us formally identify an electron state as localized if it carries
a calculated amount of Hall current less than 5 \% of the typical unit
$(-e \delta E/B)(1/L_{x})$.
Then, for this sample 55 localized states are identified, 
with 10 states of positive energy lying 
in the region $9.9 \ell < y^{\rm cm} <14.2\,\ell$ of the sample width 
and 45 states of negative energy 
lying in the region $3.2\,\ell < y^{\rm cm}<14.9\,\ell$.
[Remember that in our energy scale ``negative-energy'' states are
those that are  captured by attractive impurities.]
This contrast in both number and distribution of positive- and 
negative-energy localized states is direct evidence 
that the sample edge and disorder combine to efficiently delocalize
electrons, especially positive-energy ones. 
These localized states lie within
the energy range $-0.128\,\omega < \epsilon < 0.147\,\omega$, above
which lie only 13 edge states with $y^{\rm cm}~<~1.63\,\ell$.

\begin{figure}
\epsfxsize=8cm
\centerline{\epsfbox{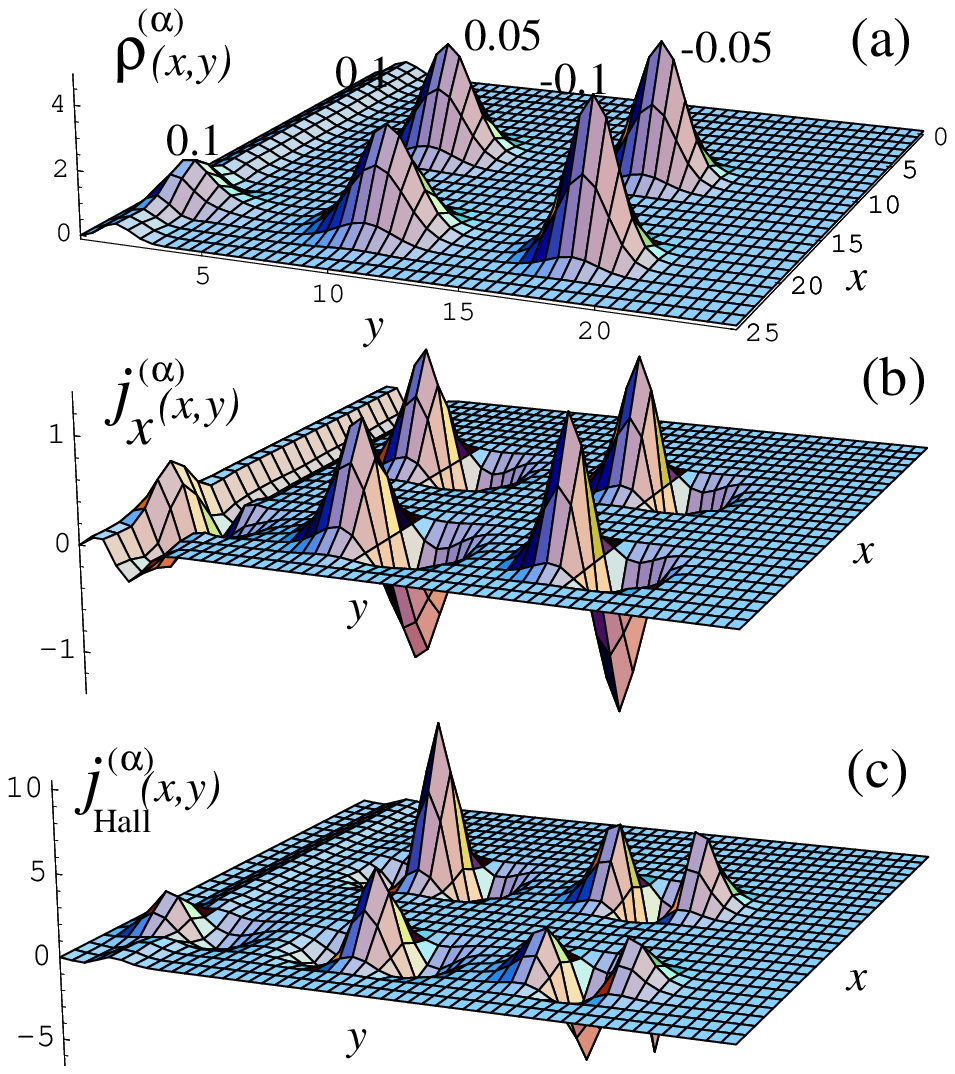}}
\caption{Electron states localized about isolated impurities and an
extended state. (a) Density distributions $\rho^{(\alpha)}(x,y)$
plotted in units of $1/(\ell L_{x})$, (b) Current distributions
$j_{x}^{(\alpha)}(x,y)$ in units of $-e\omega/L_{x}$.
(c) Hall-current distributions $j_{\rm Hall}^{(\alpha)}(x,y)$ 
in units of $-(e^{2}/2\pi \hbar)\,\delta E_{y}$.  
The coordinates $x$ and $y$ are plotted in units of the magnetic
length $\ell$.} 
\label{fig1}

\bigskip

\epsfxsize=9cm
\centerline{\epsfbox{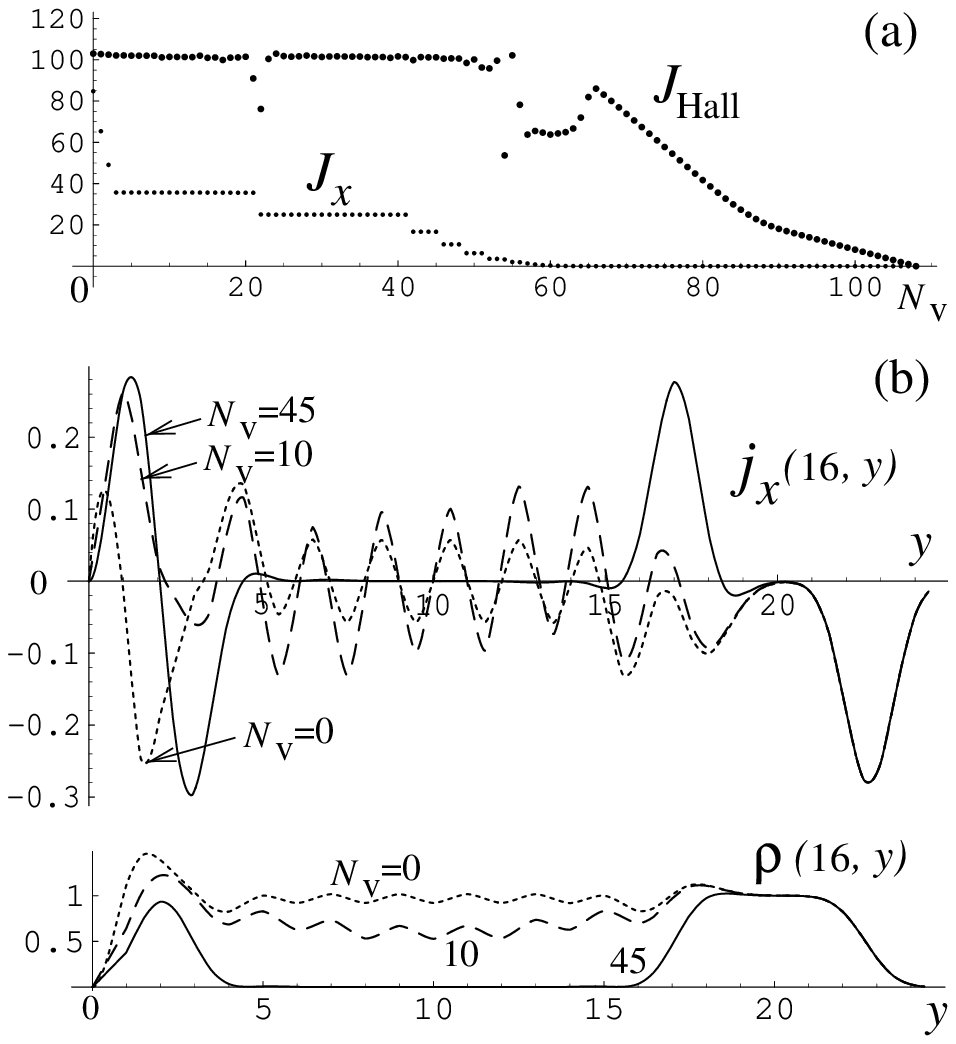}}
\vspace{0.15in}
\caption{(a) Currents vs~$N_{\rm v}$.
The total Hall current 
$J_{\rm Hall}$ is plotted in units of 
$-(e \delta E_{y}/B)(1/L_{x})$, and 
$J_{x}$ in units of $0.05\times(- e\omega \ell/L_{x})$.
(b) Current and density profiles at $x=16\ell$ 
for $N_{\rm v}=0,10,45$, with
$j_{x}(x,y)$ plotted in units of $- e \omega/(2\pi \ell)$ and 
$\rho(x,y)$  in units of $1/( 2\pi \ell^{2})$.}
\label{fig2}

\epsfxsize=8cm
\centerline{\epsfbox{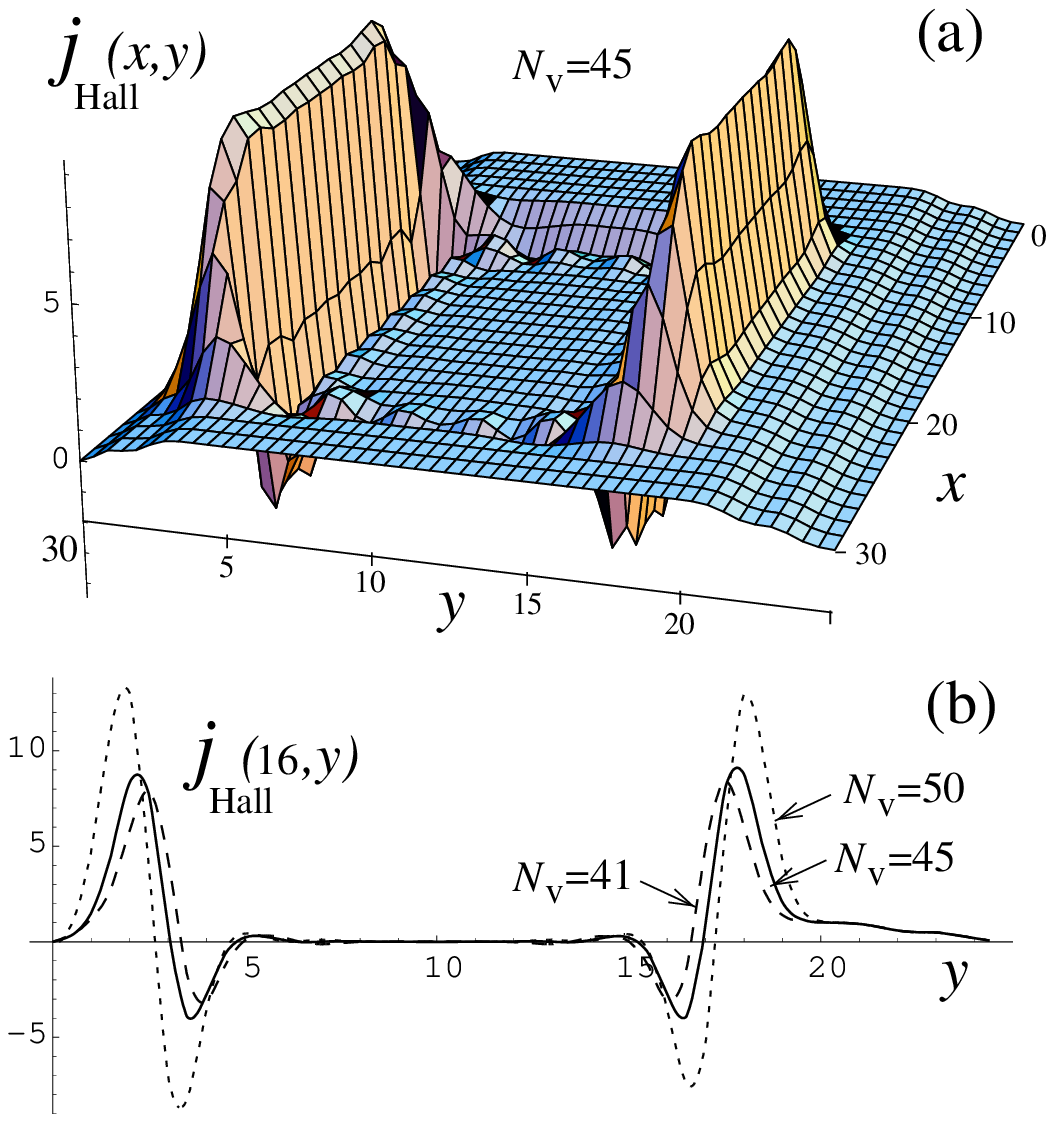}}
\vspace{0.15in}
\caption{(a) The Hall current density $j_{\rm Hall}(x,y)$ for the
$n=0$ level with $N_{\rm v}=45$, plotted in units of 
$-(e^{2}/2\pi \hbar)\,\delta E_{y}$. 
(b) Current profile at $x=16\ell$.}
\label{fig3}
  
\bigskip

\epsfxsize=9cm
\centerline{\epsfbox{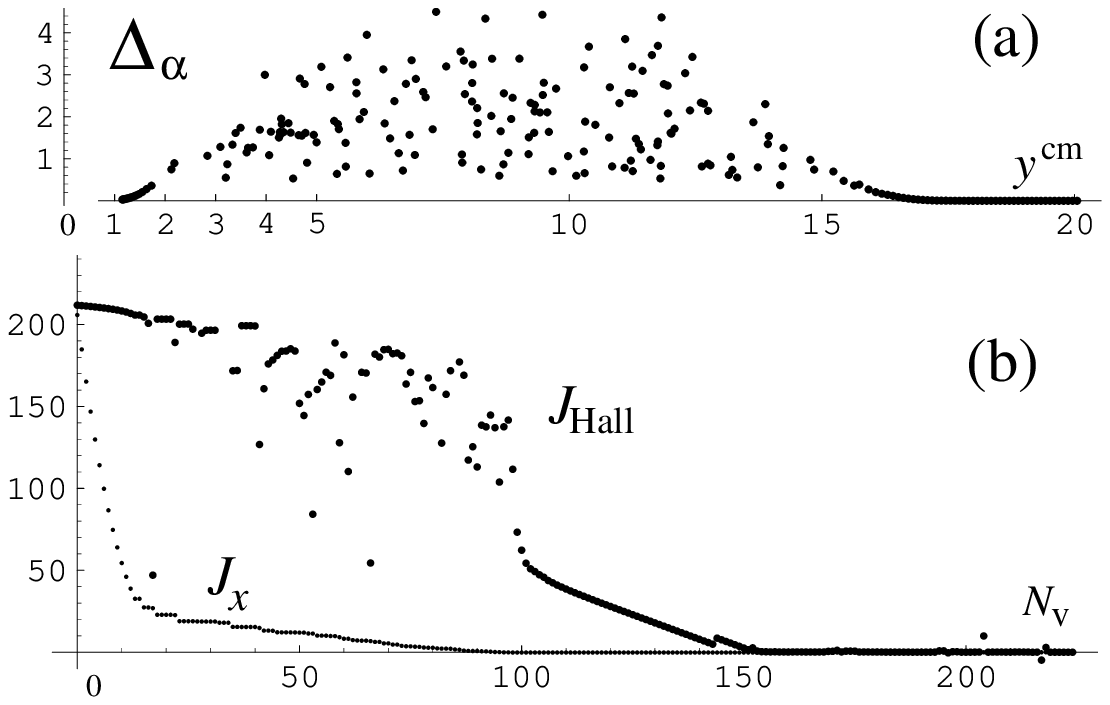}}
\caption{A sample with random impurities. 
(a) The spread $\triangle_{\alpha}$ of the wave function in
$y_{0}$ space (in units of $\ell$) plotted as a function of 
$y^{\rm cm}$ for each state. 
(b)~Currents vs~$N_{\rm v}$.  The $J_{\rm Hall}$ and $J_{x}$ are
plotted in units of $-(e \delta E_{y}/B)(1/L_{x})$ and 
$0.05 \times(- e\omega \ell/L_{x})$, respectively, with 
$\delta E_{y}/E_{y} = 0.01$ and $e\ell E_{y}/\omega = - 1/10^{5}$.}
\label{fig4}
\bigskip

\epsfxsize=9cm
\centerline{\epsfbox{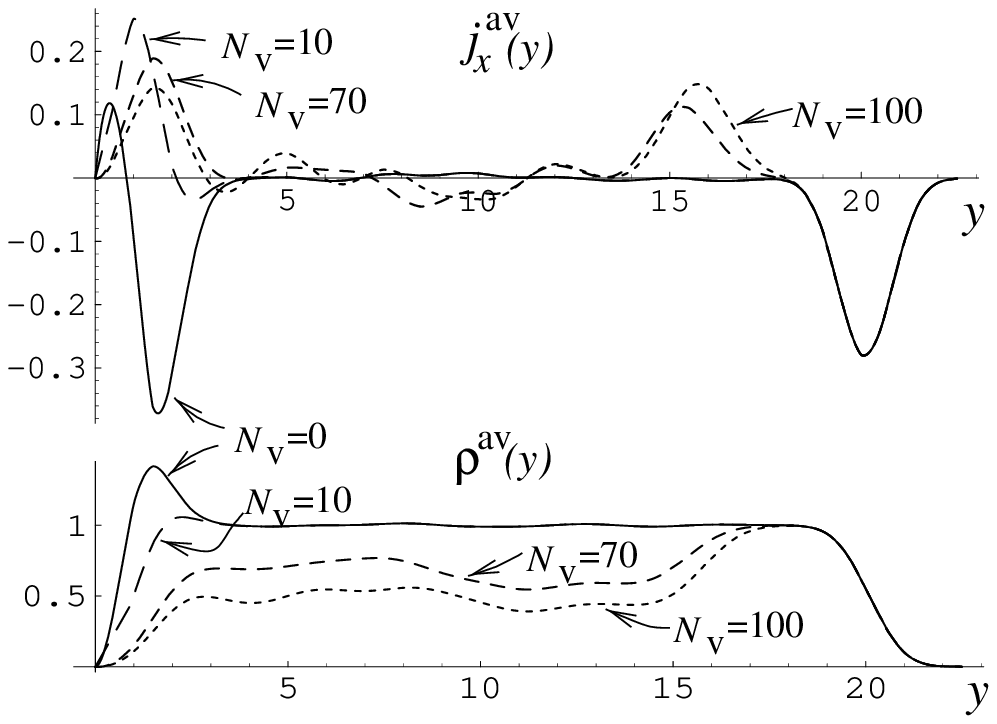}}
\caption{ Current and density distributions for the $n=0$
subband,  with $j_{x}^{\rm av}(y)$ plotted in units of 
$- e \omega/(2\pi \ell)$.}
\label{fig5}

\bigskip

\epsfxsize=8cm
\centerline{\epsfbox{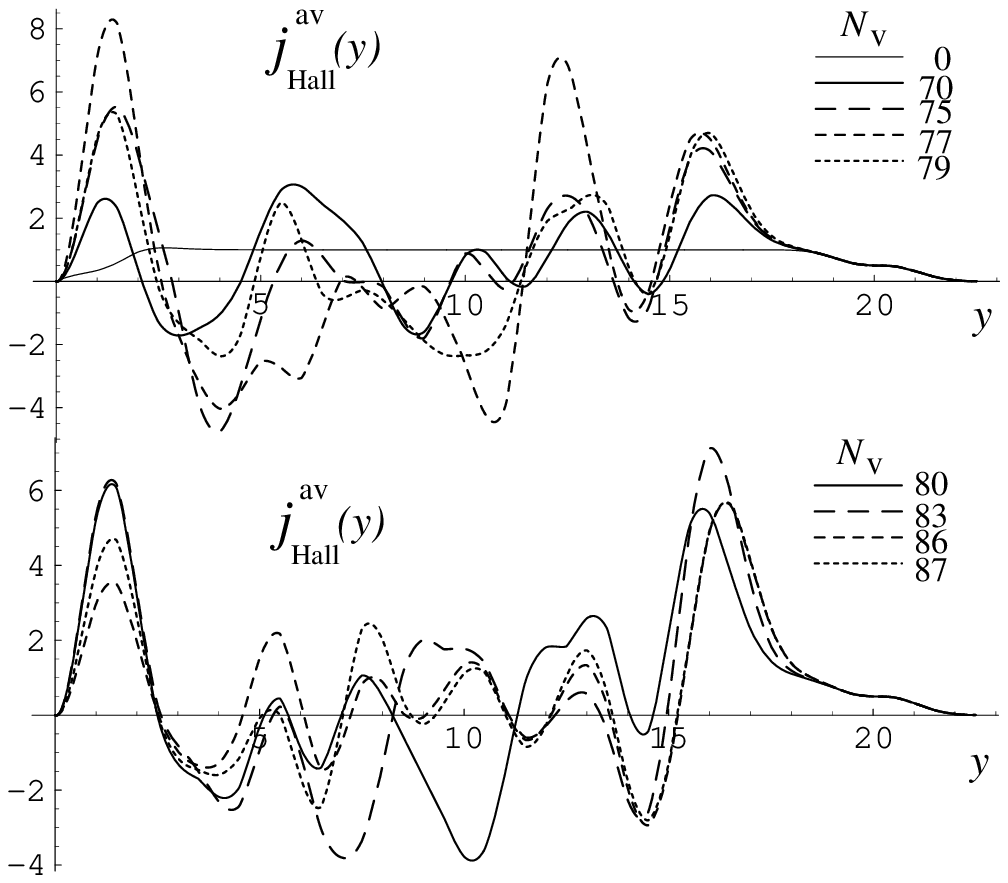}}
\caption{Hall-current distributions $j_{\rm Hall}^{\rm av}(y)$, 
in units of $-(e^{2}/2\pi \hbar)\,\delta E_{y}$, for the $n=0$
subband in the plateau regime. }
\label{fig6}
\bigskip
\newpage


\epsfxsize=7cm
\centerline{\epsfbox{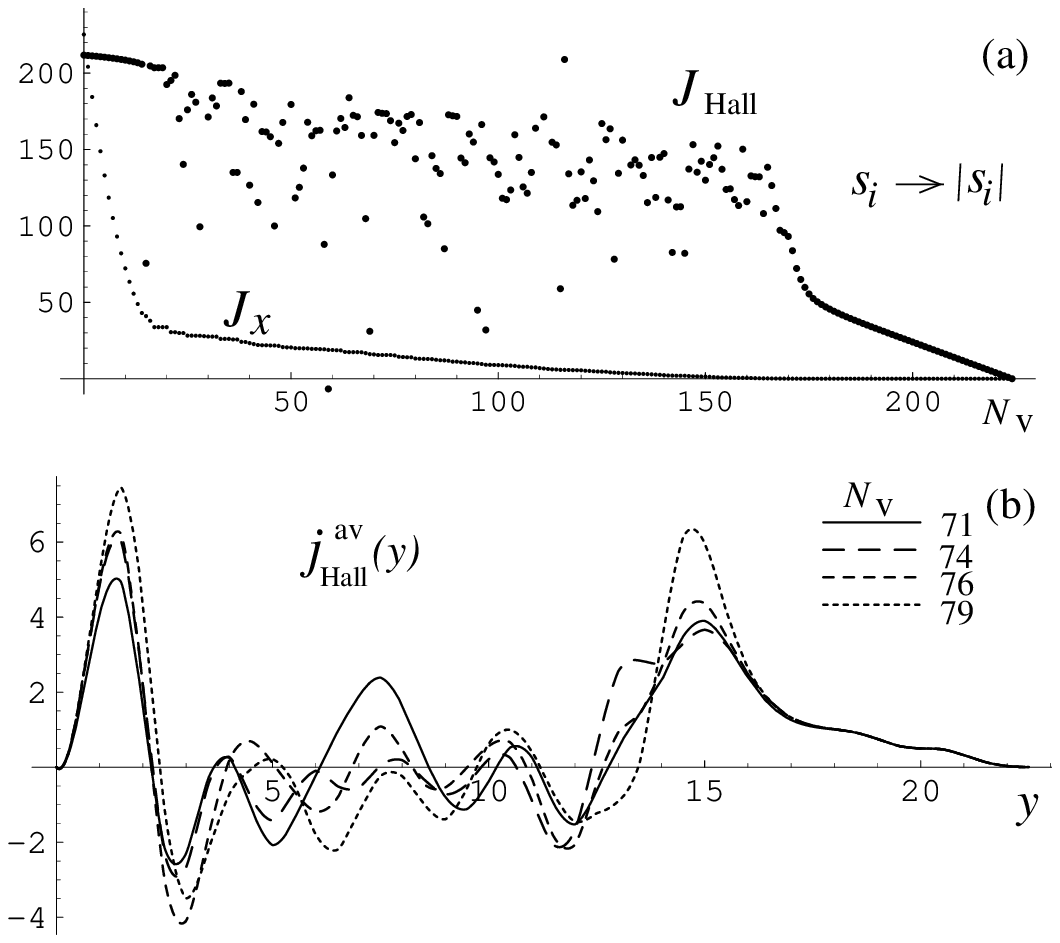}}
\caption{(a) Currents vs~$N_{\rm v}$ and 
(b) Hall-current distributions 
in the case of random repulsive impurities, 
$s_{i} \rightarrow |s_{i}|$.}
\label{fig7}
\bigskip

\epsfxsize=7cm
\centerline{\epsfbox{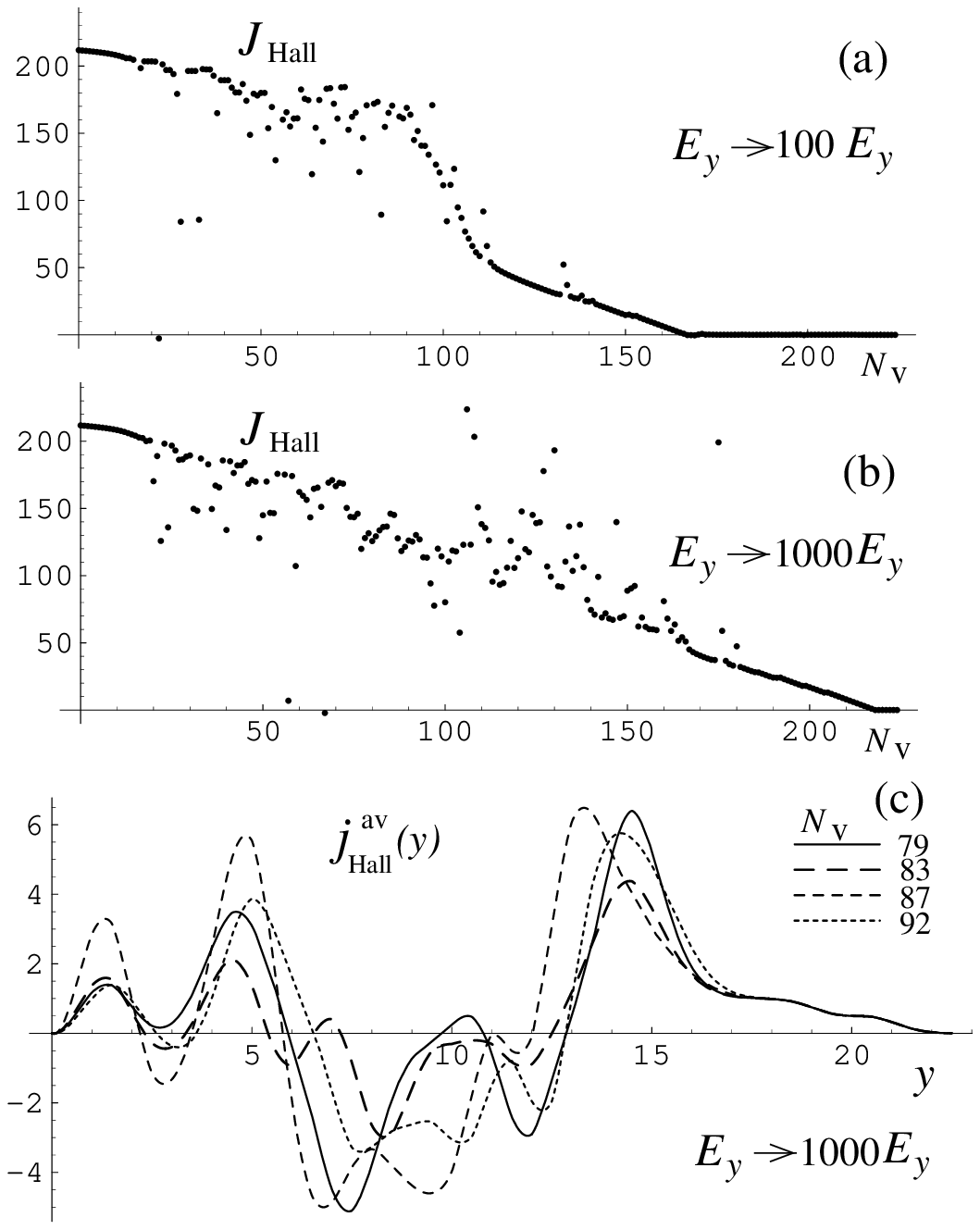}}
\caption{(a) Hall plateaus persist when $E_{y}$ is made 100 times
stronger. 
(b) Plateaus disappear when  $E_{y}$ is made 1000 times stronger.
(c) A considerable amount of Hall current thereby flows in the sample
bulk.}
\label{fig8}
\bigskip

\epsfxsize=8cm
\centerline{\epsfbox{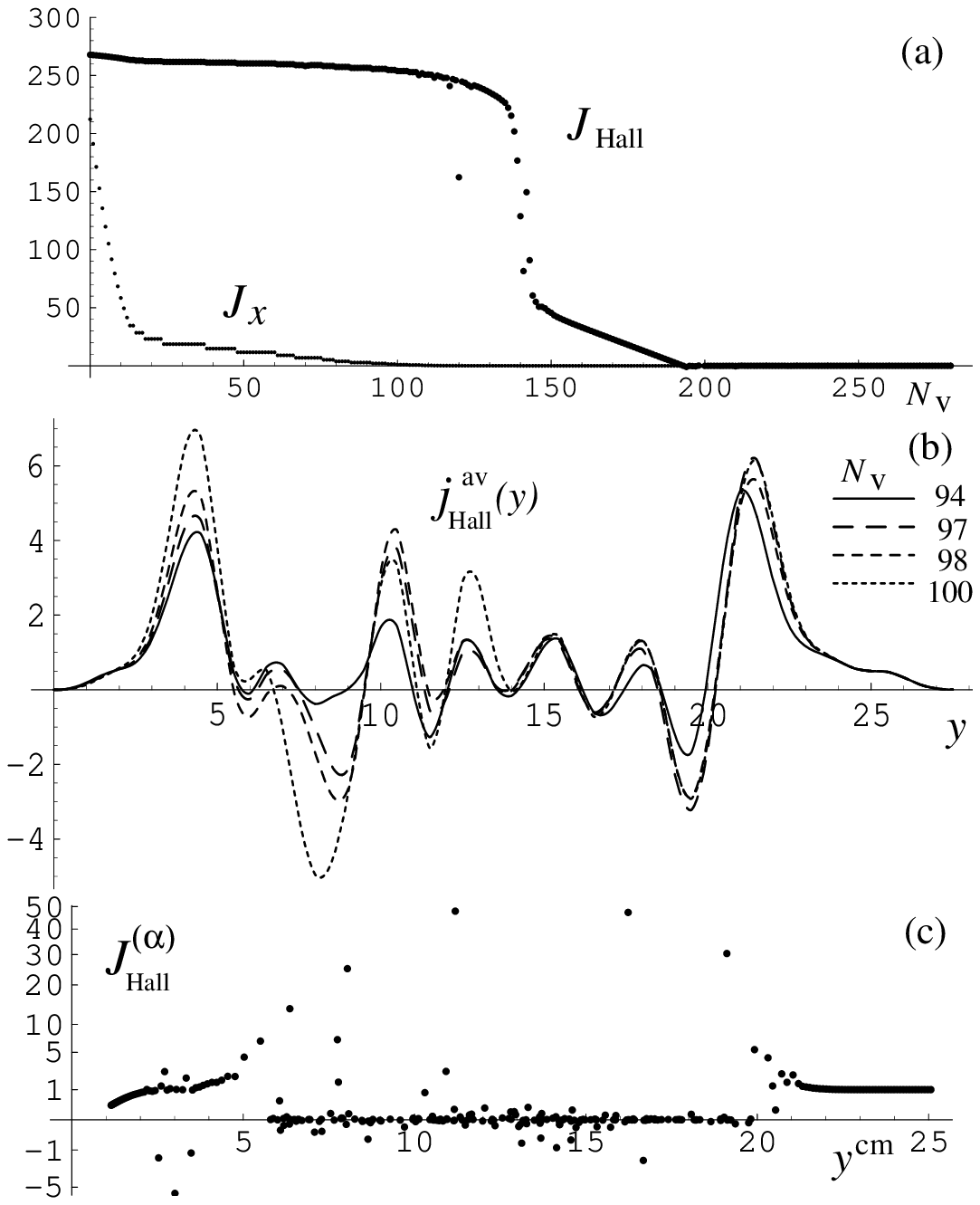}}
\caption{A Hall sample with an impurity-free buffer zone of width
$\sim 5 \ell$ along the $y=0$ edge.  
(a) Currents vs~$N_{\rm v}$. (b) Hall-current distributions. 
(c) Net amount of Hall current per state [on a square-root scale in
units of $-(e \delta E_{y}/B)(1/L_{x})$] as a function of 
$y^{\rm cm}$; here omitted for ease of exposition are 
3 pairs of points (mostly out of the range shown) that combine to 
carry only -7.8 units of current. }
\label{fig9}
\bigskip

\epsfxsize=8cm
\centerline{\epsfbox{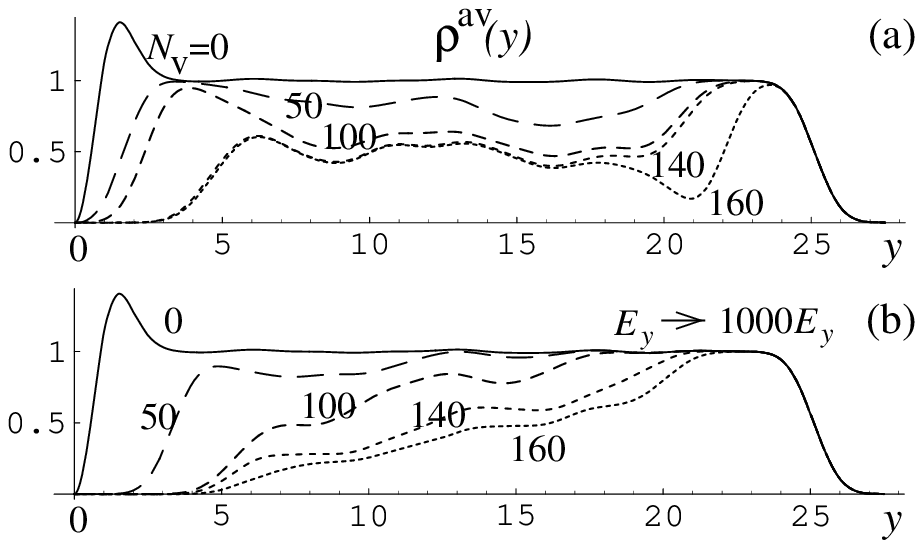}}
\caption{Density profiles for the $n=0$ subband with
$N_{\rm v}$=0,50,100,140 and 160.  
(a) $e\ell E_{y}/\omega = -1/10^{5}$ and 
(b) $e\ell E_{y}/\omega =-1/100$.}
\label{fig10}

\end{figure}

The spread in $y_{0}$ of each eigenmode $\alpha$
\begin{equation}
{\triangle}_{\alpha} = \sum_{k}|\alpha_{k}|^{2} 
|y_{0}[k] - \langle y_{0}\rangle_{\alpha}|
 \label{deltaalpha},
\end{equation}
with 
$\langle y_{0} \rangle_{\alpha} \equiv \sum_{k}|\alpha_{k}|^{2} y_{0}[k]$,
is a good measure to see the influence of disorder
on each mode.
In Fig.~\ref{fig4} (a) we plot $\triangle_{\alpha}$
as a function of $y^{\rm cm}$ for each state.
It is clear that the edge states are readily identified by their small 
spread.

Figure~\ref{fig4} (b) shows the total Hall current 
$J_{\rm Hall}$ plotted as a function of $N_{\rm v}$.
There the formation of the upper plateau for $N_{\rm v} \lesssim 90$
is incomplete because of a relatively small number of positive-energy
localized states.  A steep fall of $J_{\rm Hall}$ around 
$N_{\rm v}\sim 100$ implies that a large portion of the Hall current
is carried by the 98th$\sim$100th states residing in 
the sample bulk $8\,\ell \lesssim y^{\rm cm} \lesssim 14\,\ell$, 
each carrying $10 \sim40$ units of $(-e \delta E/B)(1/L_{x})$.
The subsequent linear decrease of $J_{\rm Hall}$
for $103 \lesssim N_{\rm v} \lesssim 153$ is due to gradual
evacuation of extended states from the impurity-free region 
$15\,\ell \lesssim y \lesssim 20\,\ell$.
Finally the negative-energy localized states give rise to a clear
lower plateau for $N_{\rm v}\ge 154$.

To examine the current distributions one has to 
recall that weaker impurities, as long as they trap electrons,
disturb $j_{\rm Hall}(x,y)$ more violently.
Correspondingly, direct inspection of $j_{\rm Hall}(x,y)$ 
does not clearly reveal the global flow of electrons.
Fortunately we find the $x$-averaged current
\begin{equation}
j_{\rm Hall}^{\rm av}(y)=(1/L_{x})\int dx\, j_{\rm Hall}(x,y)
 \label{jHav}
\end{equation}
particularly suited for this purpose.

Figure~\ref{fig5} shows the $x-$averaged current 
$j_{x}^{\rm av}(y)= (1/L_{x}) \int dx\,j_{x}(x,y)$ and density 
$\rho^{\rm av}(y)$ across the sample width. 
The paramagnetic component carried by the edge states is prominent in
$j_{x}^{\rm av}(y)$ only for $0\le N_{\rm v}\lesssim 5$.  
For $N_{\rm v}\gtrsim 10$, $j_{x}^{\rm av}(y)$ is dominated by 
the orbital diamagnetic components that flow in 
opposite directions at opposite edges. 
The current distribution formed around $y \sim 16\,\ell$ is also
a diamagnetic current associated with the extended states
surviving in the region $15\,\ell \lesssim y \lesssim
20\,\ell$, seen explicitly from $\rho^{\rm av}(y)$.
The density distribution $\rho^{\rm av}(y)$ reveals an important role
of disorder: Impurities work to distribute electrons rather evenly 
throughout the sample; in their absence  
the region $0\lesssim y \lesssim 5\,\ell$, e.g., would be 
completely evacuated for $N_{\rm v}\gtrsim 70$.

The Hall current $j_{\rm Hall}^{\rm av}(y)$ scarcely deviates from 
a uniform distribution (of the impurity-free case), 
drawn with a thin real curve in Fig.~\ref{fig6}, while only edge
states become vacant with increasing $N_{\rm v}$, i.e., for 
$0 < N_{\rm v} \lesssim 13$. 
For $N_{\rm v} \gtrsim 14$  the current distribution changes with
$N_{\rm v}$, especially in the sample bulk.
In the plateau regime $j_{\rm Hall}^{\rm av}(y)$ tends to diminish 
on average in the sample bulk $5\,\ell \lesssim y \lesssim 15\,\ell$
while forming prominent concentrations near the sample edges;
see Fig.~\ref{fig6}.

For comparison it is enlightening to look into the case where 
all the impurities are made repulsive, $s_{i} \rightarrow |s_{i}|$.
In this case no negative-energy states arise and only 3 states 
with $y^{\rm cm}> 10\,\ell$ remain localized.
No Hall plateaus are expected a priori. However, as shown in
Fig.~\ref{fig7} (a), $J_{\rm Hall}$ varies somewhat irregularly
with $N_{\rm v}$ and a vague plateau emerges as an envelope.
This implies that, while there are only few well-localized states,
a sizable fraction of electron states is nearly localized,  
each carrying a reduced amount of Hall current.
As in the original case, there is again a range of 
$N_{\rm v}$ where $j_{\rm Hall}^{\rm av}(y)$ gets small on average 
in the sample bulk; see Fig.~\ref{fig7} (b).

Let us go back to the original case again and try to vary the Hall 
field. In Fig.~\ref{fig8} (a)
we show the $J_{\rm Hall} - N_{\rm v}$ characteristics 
when $E_{y}$ is made 100 times stronger.
A comparison with Fig.~\ref{fig4} (b) reveals an apparent decrease
($\sim$ 15) in the number of the electrons forming the lower plateau;
this demonstrates that the Hall field works 
to delocalize electrons in the sample bulk. 
These liberated electrons carry a Hall current so that the steep
fall of $J_{\rm Hall}$ around  $N_{\rm v}\sim 100$ becomes somewhat 
tempered. Still no drastic changes arise in the current and density
distributions.

However, when $E_{y}$ is made 1000 times stronger, i.e.,
$-e\ell\,E_{y}/\omega = 1/100$, delocalization prevails
and only 8 states of negative energy with $y^{\rm cm}> 8.6\,\ell$
remain localized.  
An approximate linear variation of 
$J_{\rm Hall}$ with $N_{\rm v}$ in Fig.~\ref{fig8} (b)
indicates that most of the states indeed stay extended.
Still the currents are disturbed 
and there is a general tendency that $j_{\rm Hall}^{\rm av}(y)$
concentrates near the two edges, though leaving a considerable amount
in the sample bulk as well; see Fig.~\ref{fig8} (c).

Similarly, when $E_{y}$ is made 1000 times stronger, 
the vague plateau in Fig.~\ref{fig7} (a) disappears
and a sizable amount of Hall current is seen to flow in the sample
bulk. It is now clear that the decrease of the Hall current in the
sample bulk is correlated with the dominance of localized states
there. 
Figures~\ref{fig6},~\ref{fig7}(b) and~\ref{fig8}(c) 
thus give evidence that in the plateau regime the Hall current flows
with a tendency to avoid the disordered sample bulk.

To further elucidate the effect of the sample edges let us now
widen the sample width to 
$L_{y}=\sqrt{2\pi}\ell \times 10 \approx 25 \ell$, and
embed the present configuration of 180 impurities within the
region $5\,\ell \le  y \le 20\,\ell$ of the sample, leaving the two
edge regions $0 <  y < 5\,\ell$ and $20\,\ell <  y < 25\,\ell$
impurity-free.

Of the 280 electron states accommodated in the $n=0$ subband, 
we find 146 localized states lying within the same energy range 
$-0.128\,\omega < \epsilon < 0.147\,\omega$ as before, 
with 73 positive-energy states lying in the region 
$6.5\,\ell < y^{\rm cm}<19.9\,\ell$ of the sample width 
and 73 negative-energy states lying in the region 
$5.7\,\ell < y^{\rm cm} <19.8\,\ell$.
Introduction of 56 electrons in the impurity-free buffer zone of width 
$\sim 5\,\ell$ has thus brought about 91 extra localized states. 
This demonstrates again how efficiently the sharp edge and
disorder combine to delocalize electrons near the sample edges.

Since most of the electrons are localized in the the sample bulk, 
there emerge clear Hall plateaus, as shown in Fig.~\ref{fig9} (a);
there the upper plateau is not completely flat simply because the edge 
states carry a small amount of Hall current. 
Also emerges over a wide range of  $N_{\rm v}$
the tendency that the Hall current is expelled
out of the disordered region; see Fig.~\ref{fig9} (b); in general, 
as $N_{\rm v}$ approaches the mobility edge $N_{\rm v}\sim 140$,
the variation of $j_{\rm Hall}^{\rm av}(y)$ increases in magnitude.

In Fig.~\ref{fig9} (c) we plot the net amount of Hall current per
state, $J_{\rm Hall}^{(\alpha)}$, as a function of $y^{\rm cm}$ for
each state. 
It is clearly seen that electron states residing on the edges 
of the disordered region (``bulk edges'') support a considerable
amount of Hall current per state and that a small number of states in
the inner bulk carry an even larger amount.

The density distribution $\rho^{\rm av}(y)$, shown in
Fig.~\ref{fig10} (a), also reveals that in the upper-plateau regime
($N_{\rm v}\sim 100$) more electrons survive on the bulk edges than in
the sample interior.   
This feature scarcely changes even if $E_{y}$ is made 100 times
stronger.   
When  $E_{y}$ is made 1000 times stronger, impurities lose their
ability to redistribute the electrons, as seen from
Fig.~\ref{fig10} (b), and the Hall plateaus disappear.

Whether localization dominates in the sample bulk or not depends on
the competition between disorder and the Hall field.
For the cases we have discussed so far Hall plateaus disappear when 
$|s_{i}/(e\ell\,E_{y}/\omega)| \lesssim O(10)$; 
note the estimate~(\ref{svsE}) of energy cost.
For an independent check we have made all the impurities 1000 times 
weaker $s_{i}\rightarrow s_{i}/1000 \sim O(1/10^{4})$  
with $-e\ell\,E_{y}/\omega = 1/10^{5}$ kept fixed, and verified that
the plateaus in Figs.~\ref{fig4} (b),~\ref{fig7} (a) 
and~\ref{fig9} (a)  
disappear with the current distributions undergoing characteristic
changes. 
This confirms again that it is the localization of electron states 
in the sample bulk that is essential for the quantum Hall effect.

\section{Summary and discussion}
In this paper we have studied current distributions for finite Hall
samples with disorder.
The observations drawn from our numerical experiment are summarized as
follows:\hfil\break
(i) The current $j_{x}$ and the Hall current $j_{\rm Hall}$ are
substantially different in distribution.  
(ii) The fast-traveling edge states carry a large amount of current
per state but support only a negligible portion of the Hall current.
In the Hall-plateau regime the edge states (of the uppermost level) 
are vacant and scarcely affect the distributions of 
both $j_{x}$ and $j_{\rm Hall}$. The edge states are in no sense 
the principal carriers of the Hall current.
(iii) The sample edge and impurities combine to efficiently delocalize
electrons near the edge, especially those to be trapped by repulsive
impurities. 
Such extended ``bulk-edge'' states arise over a scale larger than the 
width ($\sim 3\,\ell$)  of the edge-state channels.
(iv) In the plateau regime the Hall current $j_{\rm Hall}$
tends to diminish on average in the sample bulk and concentrates 
near the sample edges. 
(v) In the transient regime between Hall plateaus the major portion of 
the Hall current is carried by a relatively small number of electron
states in the sample bulk.  
(vi) The Hall field competes with disorder to delocalize electrons in
the sample bulk. We have seen, in particular, that the breakdown of
the quantum Hall effect is caused by delocalization of electron bulk
states beyond some critical Hall field.

Observations (iii) and (iv) give evidence for the expulsion of the 
Hall current out of the disordered sample bulk, 
which is expected theoretically~\cite{KStwo} on 
the basis of exact compensation of the Hall current among extended and
localized states.  This ``bulk-edge'' Hall current is very clear for
the case of an array of impurities in Sec.~III; such an impurity array
may be realized by use of quantum dots.  
Large edge-channel widths reported experimentally~\cite{FKHBW,HTS,W}
appear to be in favor of the bulk-edge Hall current rather than 
the current carried by the edge states.

For an improved analysis the uniform field $E_{y}$ we have employed
may be replaced by a self-consistent Hall field generated internally
by an injected current.  
This would still leave all our observations (i)$\sim$(vi) essentially
intact, because the key factor there is competition of the Hall field
with disorder rather than its distribution;  
in this sense the bulk-edge channels of our concern differ
from the edge channels generated by redistributed charges.~\cite{CSG} 
Actually, the Hall field has been observed~\cite{FKHBW} to be stronger
near the sample edges than in the interior, in conformity with some of
theoretical calculations.~\cite{MRB}  
In view of observation (vi), adopting such a realistic Hall-potential
distribution will thus work to make the bulk-edge channels even wider
than in the uniform-field case.

Observation (vi) suggests that field-induced delocalization of
electrons in the sample bulk is a possible mechanism 
for the breakdown of the QHE. 
With typical values $\omega \sim$ 10 meV and $\ell \sim$ 100 \AA, 
we find a critical field of 
magnitude $|E_{y}^{\,\rm cr}|\sim$ 100 V/cm 
for the samples examined; 
in addition, keeping impurity strengths $\lambda_{i}$ fixed 
yields the power-law dependence of $E_{y}^{\,\rm cr}$ 
on the magnetic field,
$E_{y}^{\,\rm cr} \propto \ell^{-3} \propto B^{3/2}$.
A critical field of this order of magnitude and 
the $B$ dependence of this form appear consistent with
experiments.~\cite{Kawaji}  
It would be worthwhile to further explore this mechanism.

\acknowledgments

The author wishes to thank Y. Nagaoka, B. Sakita and K. Fujikawa for
useful discussions. 
This work is supported in part by a Grant-in-Aid for 
Scientific Research from the Ministry of Education of Japan, 
Science and Culture (No. 07640398).  \\

\appendix

\section{Impurity distribution}

In this appendix we display the impurity configuration used in Sec.~V.
Let us first select 45 points $\{(\bar{x}_{i},\bar{y}_{i});
i=1,\cdots,45\}$ randomly distributed
over the range $0< \bar{x} < 6$ and $0< \bar{y} < 6$.  
One such choice we adopt is\hfil\break 
P$_{45}$=\{
(0.28,~3.85), (0.34,~0.33), (0.56,~2.32), (0.40,~5.04), (0.57,~5.52),
(0.64,~1.23), (0.79,~4.78), (0.91,~5.86), (0.95,~0.61), (1.05,~3.70), 
(1.22,~2.75), (1.41,~4.43), (1.68,~3.66), (1.77,~0.73), (1.86,~0.24), 
(2.04,~5.32), (2.20,~1.62), (2.47,~2.29), (2.55,~5.46), (2.72,~2.86),  
(2.77,~1.14), (2.81,~4.46), (2.93,~3.70), (3.33,~1.83), (3.34,~4.01),  
(3.41,~2.83), (3.64,~5.15), (3.68,~0.99), (3.72,~0.26), (3.91,~3.40), 
(4.33,~4.75), (4.37,~4.19), (4.55,~1.58), (4.61,~3.32), (4.65,~0.64),  
(4.76,~4.12), (4.85,~2.88), (4.90,~5.08), (5.03,~5.48), (5.21,~1.57), 
(5.35,~3.80), (5.46,~1.11), (5.57,~2.42), (5.60,~4.66), (5.96,~0.16) \}.

Construct out of this P$_{45}$ another table 
$\{(\bar{y}_{i}+6,\bar{x}_{i}); i=1,\cdots,45\}$ (via interchange 
$\bar{x}_{i}\leftrightarrow\bar{y}_{i}$ and a shift), and
sort it into ascending order with respect to 
the $\bar{y}_{i} + 6$ components to form 
P$_{45}^{\rm t}= \{(6.16,~5.96), \cdots,  (11.86,~0.91)\}$.  
These P$_{45}$ and P$_{45}^{\rm t}$ are combined to yield a 
table of 90 points, P$_{90}= \{(0.28,~3.85),\cdots, (11.86,~0.91)\}$, 
distributed over the range $0< \bar{x} < 12$ and $0< \bar{y} < 6$.

On the other hand, we shift P$_{45}$ by (3,3) (which is our arbitrary
choice)  modulo 6 and reorder the resulting table 
$\{(\bar{x}_{i} + 3, \bar{y}_{i} + 3)$ mod 6; $i=1,\cdots,45\}$ 
into ascending order with respect to the first components to form
P'$_{45} = \{(0.33,~4.83),\cdots, (5.93,~0.7)\}$. 
We then construct out of this P'$_{45}$ a new table 
P'$_{90} = \{(0.33,~4.83), \cdots, (11.88,~1.85)\}$ of  90
points by the same procedure as done for P$_{90}$ out of P$_{45}$.

Finally we shift P'$_{90}$ by (12,0),
$(\bar{x}_{i},\bar{y}_{i}) 
\rightarrow (\bar{x}_{i}+ 12,\bar{y}_{i})$, and add it to  P$_{90}$, 
obtaining the table of 180 points
P$_{180} = \{(0.28,~3.85), \cdots,(23.86,~5.72), (23.88,~1.85)\}$
distributed over the range $0< \bar{x} < 24$ and $0< \bar{y} < 6$.   
Via the rescaling $x_{i}=\sqrt{2\pi}\ell \times (28/24)
\times \bar{x}_{i}$ and $y_{i}=\sqrt{2\pi}\ell \times \bar{y}_{i}$ 
we distribute these 180 impurities at positions $(x_{i},y_{i})$ within 
the region $0\le x \lesssim 70\,\ell$ and $0\le y \lesssim 15\,\ell$
of the sample.

The strengths $s_{i}$ we adopt for the first 45 impurities are\hfil\break
S$_{45} = 0.1 \times$ \{-0.973, -0.826, 0.855, -0.368, 0.793, 
  0.316, -0.233, 0.180, -0.692, 0.916, 
  -0.463, 0.252, -0.452, -0.745, -0.485, 
  0.446, -0.622, -0.076, 0.503, 0.205, 
  -0.973, 0.758, 0.544, 0.720, 0.997, 
  0.684, 0.689, 0.088, -0.874, -0.631, 
  -0.078, 0.908, 0.818, -0.548, -0.6141, 
  -0.344, 0.271, -0.802, 0.871, 0.218, 
  -0.106, 0.283, -0.632, -0.986, -0.132\},\hfil\break
which are generated randomly within the range $|s_{i}| \le 0.1$.
To generate 180 strengths we arbitrarily choose to set 
$s_{i}=(0.1-|s_{i-45}|)\, {\rm sign}(s_{i-45})$ for $46 \le i \le 90$ 
and then $s_{i}=- s_{181-i}$ for $91 \le i \le 180$.


\newpage

\begin{thebibliography}{99}
\bibitem{fn}  Inter-subband degeneracy is properly taken care of if one
diagonalizes the Hamiltonian in full $(n,y_{0})$ space 
(or its truncation with a finite number of subbands retained).
\end{thebibliography}
\end{document}